\documentclass[twocolumn,preprintnumbers,superscriptaddress,amsmath,amssymb]{revtex4}
\usepackage{mathrsfs}
\usepackage{graphicx}

\begin{document}

\title{Co-induced nano-structures on Si(111) surface\cite{note}}

\author{Y. T. Cui}
\email[Corresponding author:]{yitaocui@gmail.com}
\affiliation{Hiroshima Synchrotron Radiation Center, Hiroshima
University, Higashi-Hiroshima 739-0046 Japan}

\author{T. Xie}
\affiliation{Hiroshima Synchrotron Radiation Center, Hiroshima
University, Higashi-Hiroshima 739-0046 Japan}

\author{M. Ye}
 \affiliation{Graduate School of Science, Hiroshima University
Higashi-Hiroshima 739-8526 Japan}

\author{A. Kimura}
\email[Corresponding author:]{akiok@hiroshima-u.ac.jp}
\affiliation{Graduate School of Science, Hiroshima University
Higashi-Hiroshima 739-8526 Japan}

\author{S. Qiao}
\email[Corresponding author:]{qiaoshan@fudan.edu.cn}
\affiliation{Hiroshima Synchrotron Radiation Center, Hiroshima
University, Higashi-Hiroshima 739-0046, Japan} \affiliation{Advanced
Materials Laboratory, Physics Department and Surface Physics
Laboratory (National Key Laboratory), Fudan University, Shanghai
200433, China}

\author{H. Namatame}
\affiliation{Hiroshima Synchrotron Radiation Center, Hiroshima
University, Higashi-Hiroshima 739-0046, Japan}

\author{M. Taniguchi}
 \affiliation{Graduate
School of Science, Hiroshima University Higashi-Hiroshima 739-8526
Japan} \affiliation{Hiroshima Synchrotron Radiation Center,
Hiroshima University, Higashi-Hiroshima 739-0046, Japan}

\begin{abstract}

The interaction of cobalt atoms with silicon (111) surface has been
investigated by means of scanning tunneling microscopy (STM) and
low-energy electron diffraction (LEED). Besides the Co silicide
islands, we have successfully distinguished two inequivalent
Co-induced $\sqrt{13}\times\sqrt{13}$ reconstructions on Si (111)
surface. Our high-resolution STM images provide some structural
properties of the two different $\sqrt{13}\times\sqrt{13}$ derived
phases. Both of the two phases seem to form islands with single
domain. The new findings will help us to understand the early stage
of Co silicide formations.

\end{abstract}


\maketitle

\section{Introduction}
Transition-metal silicides have been widely investigated in recent
years in terms of microelectronic industry and fundamental science.
Besides the applications of transition-metal silicides (such as
CoSi$_2$, NiSi$_2$) in integrated-circuits, the effects of adsorbed
atoms on a reconstructed Si surface and the mechanisms and
coefficients of surface diffusion are also fundamental subjects of
surface science. Although there have been extensive studies on the
growth and properties of CoSi$_2$ epitaxial films, knowledge about
the structural details of Co adatoms and the initial stages of
cobalt disilicide are insufficient partly due to an appearance of
very rich reconstructed phases.
Si(111)-$\sqrt{7}$$\times$$\sqrt{7}$R19.1$^\circ$-Co (hereafter as
$\sqrt{7}$) structure has been studied by means of LEED, XPS (X-ray
Photoemission Spectroscopy), ARPES (Angle-Resolved Photoelectron
Spectroscopy), STM and LDA (Local Density Approximation) based
calculations \cite{Pirri29,Bennett69,Bennett48}. Two types of
domains with $\sqrt{13}$$\times$$\sqrt{13}$R13.9$^\circ$ (hereafter
as $\sqrt{13}$) reconstructions and both the $\sqrt{13}$ and
$\sqrt{19}$$\times$$\sqrt{19}$R23.4$^\circ$ (hereafter as
$\sqrt{19}$) reconstructions were found by LEED and STM
 \cite{Dolbak373,Loffler81}, but the detailed structural information
is still lacking. In order to investigate the detailed surface
structures, STM experiment was carried out. In this paper, we will
show two inequivalent Co-induced $\sqrt{13}$ structural phases
through several real space STM images together with their FFT (fast
Fourier transformation) and LEED patterns.

\section{Experimental}

The Co induced surface structures were fabricated and studied with a
commercial room-temperature scanning tunneling microscope (STM, RHK
UHV-400) mounted in an ultrahigh-vacuum (UHV) chamber combined with
a sample-preparation chamber. The base pressures of the two chambers
were around 1.0$\times$10$^{-10}$ Torr. Clean Si(111) (\emph{p}
type, $\geq$ 1000 $\Omega$$\cdot$cm) surfaces were obtained by the
following process. After degassed for several hours at 800 K, the
sample was flashed several times to 1470 K for 5 seconds, then the
sample temperature was rapidly reduced to about 1170 K, and slowly
decreased to room temperature at a pace of 1$\sim$2 K/s with a
pressure below 1.0$\times$10$^{-9}$ Torr. The sample temperature was
controlled by direct-current-heating and a nearly perfect 7$\times$7
reconstruction was obtained by this method. Co was deposited from a
water-cooled electron-beam evaporator (Omicron EFM 3) to clean
7$\times$7-Si(111) surface with a rate of $\sim$0.08 ML/min (where 1
ML = 7.83$\times$10$^{14}$ atoms/cm$^2$). The temperature of silicon
substrate was kept at 800 K during the Co deposition. Auger electron
spectroscopy (AES) and LEED observations were routinely carried out
to check the surface cleanliness and reconstruction. All of the STM
images were taken at room temperature.

\section{Results and discussion}

\begin{figure}
\centering
\includegraphics[width=6cm]{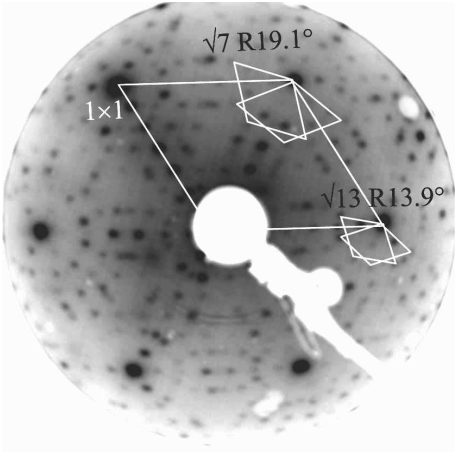}
\caption{\label{fig:1} LEED pattern with electron energy of 63.4 eV.
The reciprocal unit cells of Si(111)-1$\times$1, $\sqrt{13}$ and
$\sqrt{7}$ reconstructed surfaces are marked.}
\end{figure}

The LEED patten of 0.5 ML Co on Si(111) surface is presented in
Fig.~\ref{fig:1}. Besides 1$\times$1, 7$\times$7 and $\sqrt{7}$
patterns, one can find the spots of $\sqrt{13}$ derived structure,
which are the same as those in the published
results\cite{Dolbak373}. It should be mentioned here that these
additional spots were formerly attributed to two types of domains
with $\sqrt{13}$ phase by Dolbak et al \cite{Dolbak373}, which were
defined as A and B domains in their paper. The two kinds of domains
are separated with respect to Si (1$\bar{1}$0) mirror-image symmetry
plane and they widely exist in $\sqrt{n}$ structures such as
Ni-induced $\sqrt{19}$ twinned atomic structures\cite{Kinoda461},
Si(111)-$\sqrt{21}$$\times$$\sqrt{21}$-Ag surface with different
orientations of [11$\bar{2}$]$\pm$10.89$^\circ$ at OPB3
(out-of-phase-boundary of the $\sqrt{3}$-Ag phase)\cite{Tong64} and
so on. The present LEED pattern also indicates two domains for the
$\sqrt{7}$ surface phase as shown in Fig.~\ref{fig:1}.

\begin{figure}
  \centering
  \includegraphics[width=6cm]{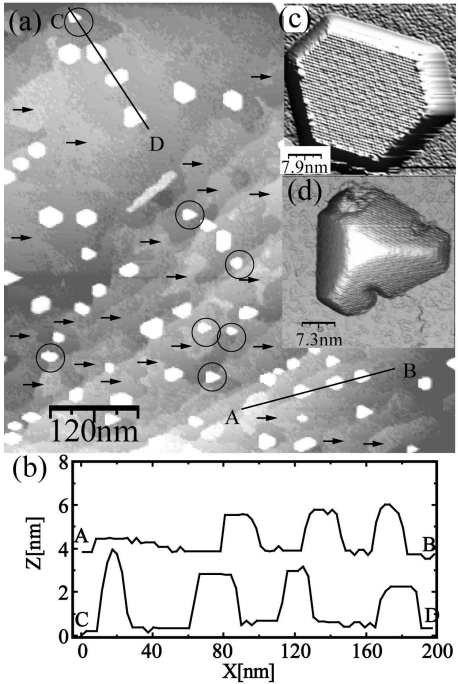}
  \caption{\label{fig:2} Typical occupied-state STM images ($V$$_S$ =
  -2.5 V, $I$$_S$ = 0.51 nA) of Si (111) surface after deposition of 0.5
  ML Co at a temperature of 800 K; (a) Typical
  topographic image in the wide area (600nm$\times$600nm); (b) The
  cross section profiles  along the lines shown in (a);  Close view of typical Co
  islands in three dimensional form: (c) atomically flat top island with 2$\times$2 structure and (d) pyramid-like island.}
\end{figure}

Fig.~\ref{fig:2} (a) shows the typical occupied-state STM image in
the wide area (600nm$ \times$600nm, $V$$_S$ = -2.5 V, $I$$_S$ = 0.51
nA).  Here, one can find some islands with larger height (brighter
image). The cross sections along two lines marked on
Fig.~\ref{fig:2} (a) are shown in Fig.~\ref{fig:2} (b). The lateral
size of islands is in the order of tens nanometers, and mostly in
the range between 20 to 50 nm with the height of about 2$\sim$5 nm.
These islands can be classified into two typical types. The major
one shows approximately triangular, trapezoidal or hexagonal shape
with atomically flat top that shows well-known
CoSi$_2$(111)-(2$\times$2) structure \cite{Bennett11,Zilani}. A
typical image of this kind of island is shown in Fig.~\ref{fig:2}
(c) in three dimensional form. The minority one as marked by
ellipses in Fig.~\ref{fig:2} (a) shows pyramid-like feature without
any CoSi$_2$(111)-(2$\times$2) structure on the triangular flat top
as seen in Fig.~\ref{fig:2} (d). It is a little higher than the
former type. The edges of both two types of islands are aligned
parallel to those of the 7$\times$7 unit cell, and the lengths are
integer times as large as that of 7$\times$7 unit cell as have been
discussed elsewhere\cite{Bennett11}. Due to the small percentage of
island relative to the whole surface, the structures of the island
surface did not reflect clear corresponding spots in LEED pattern.
It means that the surface structures shown in LEED pattern mainly
come from the contribution of inter island.

\begin{figure}
\centering
\includegraphics[width=6.5cm]{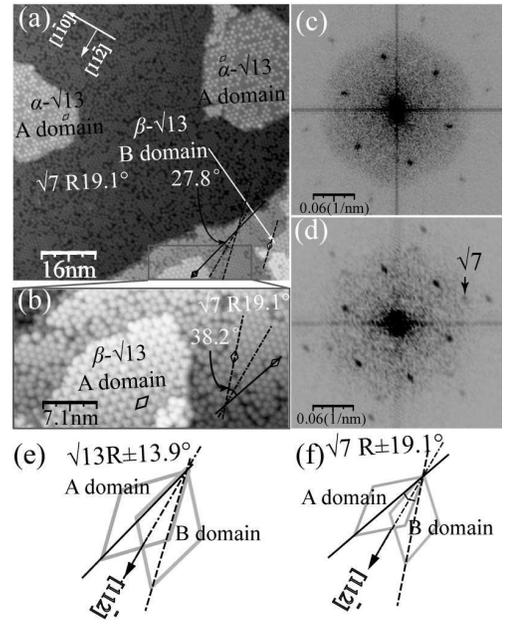}
\caption{\label{fig:3}(a) Occupied-state STM images of
$\alpha$-$\sqrt{13}$ phase with A type domain and
$\beta$-$\sqrt{13}$ phase with both A and B domains (80 nm$\times$80
nm). (b) Close-up view of $\beta$-$\sqrt{13}$ phase with B type
domain (35.5nm$\times$18nm, $V$$_S$ = -2.1 V, $I$$_S$ = 0.3 nA). (c)
FFT pattern of Fig. (a). (d) FFT pattern of Fig.~\ref{fig:3} (b)
(The $\sqrt{7}$ FFT spot is marked with an arrowhead \cite{note1}).
Schematic diagrams of the relationships of two (A and B) domains for
(e) $\sqrt{13}$ and (f) $\sqrt{7}$ surface phases.}
\end{figure}

In the inter-island regions, besides 7$\times$7, Co-induced
$\sqrt{7}$ and $\sqrt{19}$ phases, one can find irregular flat areas
with a height of about 0.3 nm (some typical regions were marked by
black arrowheads in Fig.~\ref{fig:2} (a)). More details of the
occupied-state STM images in this region are shown in
Figs.~\ref{fig:3} (a) and (b). Here, two different structures are
found. One shows a $\sqrt{13}$ surface phase, which has been
observed by L\"offler et al \cite{Loffler81} (denoted as $\alpha$
phase with A type domain). It is found that the length of unit cell
is 1.38 nm and its orientation is rotated by an angle of
13.9$^\circ$ relative to [11$\bar{2}$] direction in clockwise
direction as illustrated in Fig.~\ref{fig:3} (e). Another structure
in Fig.~\ref{fig:3} (a) and magnified in Fig.~\ref{fig:3} (b) has
not been observed before. In order to know the structural property,
we have analyzed the occupied-state STM image carefully. It is found
that the unit cell size is around 1.38 nm and the unit cell is
rotated by an angle of 13.9$^\circ$ in clockwise direction relative
to [11$\bar{2}$] direction (illustrated in Fig.~\ref{fig:3} (e))
indicating the same periodicity of $\sqrt{13}$ in this new surface
phase. To distinguish with $\alpha$-$\sqrt{13}$ phase, this new
phase is denoted as $\beta$ phase with A type domain. The fast
Fourier transformation (FFT) images of Figs.~\ref{fig:3} (a) and (b)
using WSXM software\cite{Horcas78} are shown in Figs.~\ref{fig:3}
(c) and (d), respectively. The sharp FFT patterns of $\sqrt{13}$
surface phases can be seen well \cite{note1}, and both of them show
the same orientation, which together with their corresponding STM
images gives the strong evidences of the two inequivalent
$\sqrt{13}$ surface phases.

It has been mentioned above that most $\sqrt{n}$ surface phases on
Si (111) surface have two domains with a misfit angle owing to Si
(1$\bar{1}$0) mirror-image symmetry planes shown in LEED pattern.
For easily understood the different domains the schematic diagrams
of the relationship of the different domains (here, A and B domains)
for $\sqrt{13}$ and $\sqrt{7}$ reconstructions are shown in
Figs.~\ref{fig:3} (e) and (f). So, it is reasonable to observe both
domains of the two $\sqrt{13}$ phases on the surface (not shown
here). It should be mentioned that both $\alpha$ and $\beta$ phases
seem to form islands with one domain, which is not like $\sqrt{7}$
and $\sqrt{19}$ structures that form twined structures with defects.
The defect structures will restrain themselves from forming large
islands. This feature indicates that the defect-free, single-domain
$\sqrt{13}$ island-like structures may be one kind of early-stage
formation of Co silicide islands on Si (111) surface. From this
point of view, the investigation of $\sqrt{13}$ structures will help
us to understand the early stage of Co silicide formations.

\begin{figure}
\centering
\includegraphics[width=7cm]{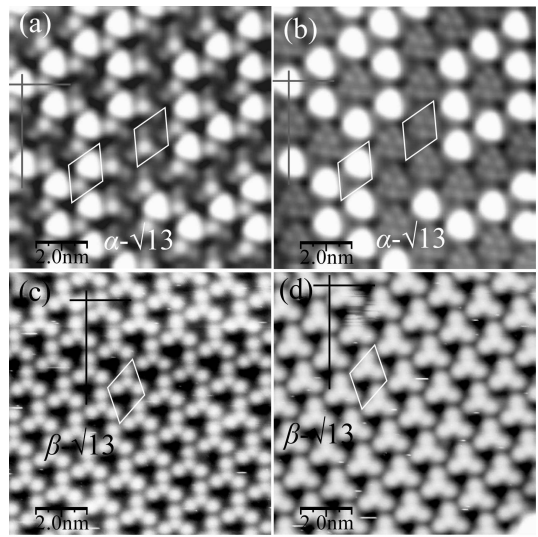}
\caption{\label{fig:4} High-resolution empty-(left) and
occupied-(right) state STM images of $\alpha$- (up case) and
$\beta$- (lower case) $\sqrt{13}$ phases (10 nm$\times$10 nm)
$V$$_S$ = $\pm$1.5 V, $I$$_S$ = 0.3 nA for (a) and (b); $V$$_S$ =
$\pm$2.0 V, $I$$_S$ = 0.8 nA for (c) and (d). The crossehairs are
located in identical positions in both of the occupied- and
empty-state images.}
\end{figure}

In order to study the detailed structures of $\sqrt{13}$ surface
phases, high-resolution STM observations have been performed. The
empty- and occupied-state STM images of
 $\alpha$- (upper) and $\beta$- (lower) $\sqrt{13}$ phases
 are presented in Fig.~\ref{fig:4} with
the unit cell indicated in the figures. The prominent feature of
both $\sqrt{13}$ surface phases is the centered hexagonal array of
threefold symmetric protrusions as is evident in Fig.~\ref{fig:4}.
Note that the $\alpha$ phase seems to comprise two layers of
triangular clusters on the top and another triangle with six
protrusions below them as shown in the occupied-state STM image
(Fig.~\ref{fig:4} (b)). Both of them just occupies half of the unit
cell (here after as HUC), the other half is a dark hole at negative
sample bias( shown in Fig.~\ref{fig:4} (b)). For the lower layer, at
a negative sample bias, the triangle with six protrusions and the
other dark HUC can be observed, while at positive sample bias the
dark HUC becomes brighter and the number of protrusions is reduced
to three at the corner with a dark hole in the center. For the upper
layer, it is noticed that the bright triangle always occupies the
same HUC -- on the top of six-protrusion triangle in occupied-states
-- and the positions of the bright triangle are not changed at
various sample bias, which means that the protrusions of the bright
triangle give the sites information of real atoms.

The expanded high-resolution $\beta$ phase images show four
protrusions (tetramers) and a dark HUC in both empty- and
occupied-states. In contrast to the $\alpha$ phase, the $\beta$
phase exhibits a little difference in spatial distribution between
empty- and occupied-state, except the protrusions in occupied-state
STM images seem fatter than that in empty-state STM images. It
should be mentioned that we still can not give the information of Co
atoms contribution in $\alpha$- and $\beta$- $\sqrt{13}$ surface
phases just from STM image. For more information, deeper
investigation and a atomic structural model need to be made and some
calculated results should be helpful.

\section{Conclusion}

In summary, self-assembled Co silicide have been fabricated on
Si(111) and investigated by means of scanning tunneling microscopy
(STM) and low-energy diffraction (LEED). The LEED pattern has
verified the existence of $\sqrt{13}$ phases on Si(111) surface. Our
high resolution STM images show that besides the Co silicide island,
there are two inequivalent $\sqrt{13}$ surface phases in the
inter-island regions. Both of the two phases seem to form islands
with single domain. From this point of view, the new findings will
help us to understand the early stage of Co silicide formations.

This work was supported by the Ministry of Education, Culture,
Sports, Sciences and Technology of Japan.


\begin{thebibliography}{99}

\bibitem{note}
Published in \emph{Applied surface science} \textbf{254} (2008)
7684-7687 doi:10.1016/j.apsusc.2008.01.136

\bibitem{Pirri29}
 C. Pirri, J. C. Peruchetti, G. Gewinner and J. Derrien, Phys. Rev. B \textbf{29} (1984) 3391; B \textbf{30} (1984) 6227.

\bibitem{Bennett69}
P. A. Bennett, M. Copel, D. Cahill, J. Falta and R. M. Tromp, Phys.
Rev. lett. \textbf{69} (1992) 1224.

\bibitem{Bennett48}
 M. -H. Tsai, J. D. Dow, P. A. Bennett, D. G. Cahill, Phys. Rev. B \textbf{48} (1993) 2486.

\bibitem{Dolbak373}
A. E. Dolbak, B. Z. Olshanetsky, S. A. Teys, Surf. Sci. \textbf{373}
(1997) 43.

\bibitem{Loffler81}
M. L\"offler, J. Cord\'on, M. Weinelt, J. E. Ortega, T. Fauster,
Appl. Phys. A \textbf{81} (2005) 1651.

\bibitem{Kinoda461}
G. Kinoda, K. Ogawa, Surf. Sci. \textbf{461} (2000) 67.

\bibitem{Tong64}
 Xiao Tong, Satoru Ohuchi, Norio Sato, Takehiro Tanikawa, Tadaaki Nagao, Iwao Matsuda, Yoshinobu Aoyagi,
and Shuji Hasegawa, Phys. Rev. B \textbf{64} (2001) 205316

\bibitem{Bennett11}
P. A. Bennett, S. A. Parikh, D. G. Cahill, J. Vac. Sci. Technol. A
\textbf{11} (1993) 1680.

\bibitem{Zilani}
M. A. K. Zilani, Lei Liu, H. Xu, Y. P. Feng, X. -S. Wang and A. T.
S. Wee, J. Phys.: Condens. Matter \textbf{18} (2006) 6987


\bibitem{Horcas78}
I. Horcas, R. Ferm\'andez, J. M. G\'omez-Rodr\'iguez, J. Colchero,
J. G\'omez-Herrero, A. M. Baro, Rev. Sci. Instrum. \textbf{78}
(2007) 013705.

\bibitem{note1}
One can see a large number of $\sqrt{7}$ reconstructions from the
STM images (See Figs.~\ref{fig:3} (a) and (b)). Since there are many
defects in $\sqrt{7}$ phases, one can not get very sharp FFT spots
with $\sqrt{7}$ structure. The irregular lattices just contribute to
a dark background as shown in Figs.~\ref{fig:3} (c) and (d).

\end{thebibliography}
\end{document}